\documentclass[prl,twocolumn,amsmath,amssymb,nofootinbib,balancelastpage,superscriptaddress]{revtex4}
\usepackage{multirow}
\usepackage{epsfig}
\usepackage{amsmath}
\usepackage{bm}
\usepackage{times}
\usepackage{graphicx}
\usepackage{color}
\usepackage{slashed}
\usepackage{graphicx}
\usepackage{amsmath}
\usepackage{relsize} 
\usepackage[latin1]{inputenc}

\usepackage[
	colorlinks=true,
	citecolor=black,
	linkcolor=black,
	urlcolor=blue,
	hypertexnames=false]{hyperref}

\def\bea{\begin{eqnarray}}
\def\eea{\end{eqnarray}}
\def\bean{\begin{equation*}}
\def\eean{\end{equation*}}
\newcommand{\arXiv}[2]{\href{http://arxiv.org/pdf/#1}{{\tt #2/#1}}}
\newcommand{\arXivold}[1]{\href{http://arxiv.org/pdf/#1}{{\tt #1}}}

\begin{document}

\title{Twisted Sisters: KK Monopoles and their Zero Modes}

\author{Csaba Cs\'aki}
\affiliation{Department of Physics, LEPP, Cornell University, Ithaca, NY 14853}
\author{Yuri~Shirman}
\affiliation{Department of Physics and Astronomy, University of California, Irvine, CA 92697}
\author{John~Terning}
\affiliation{Department of Physics, University of California, Davis, CA 95616}
\author{Michael~Waterbury}
\affiliation{Department of Physics and Astronomy, University of California, Irvine, CA 92697}

\begin{abstract}
We find the conditions for the existence of fermionic zero modes of the fundamental representation in the background of a Kaluza-Klein (KK) monopole. We show that while there is no zero mode without a real mass, a normalizable zero mode appears once the real mass is sufficiently large. This provides an elegant explanation for the known decoupling of KK monopole effects in supersymmetric theories when a large real mass term is added. We also present an application where the correct counting of KK zero modes plays an essential role in understanding the non-perturbative effects  determining the low-energy dynamics. 
\end{abstract}

\maketitle

\section{Introduction}

Magnetic monopoles can have important dynamical and phenomenological consequences under a wide variety of circumstances. They can appear as massive extended objects in grand unified theories (GUTs), their existence can explain charge quantization, or they can be responsible for non-perturbative effects in strongly coupled gauge theories. Most of the recent advances in understanding the dynamics of monopoles arose in supersymmetric gauge theories, where monopoles can become light under controlled circumstances, as for example in the famous Seiberg-Witten solution to ${\mathcal N}=2$ SUSY gauge theories \cite{Seiberg:1994rs}. Monopoles can arise as static 4D objects, but if the dimension along their world-line is compactified then the monopoles  play the role of instantons in the resulting 3D theory. These so-called monopole-instantons are essential for understanding the dynamics of 3D SUSY gauge theories, as well as theories on ${\bf R}^3\times {\bf S}^1$, which can interpolate between the 4D and the 3D theories~\cite{Seiberg:1996nz,Aharony:1997bx,deBoer,MithatEric,Aharony:2013dha,Csaki:2014cwa}. In fact, many of the non-perturbative effects in the 4D SUSY gauge theories can be best understood by considering the theory on the circle, and eventually taking the infinitely large radius limit\footnote{Monopole condensation can be used to break supersymmetry \cite{Csaki:2011nw}, while monopole contributions can sometimes be under control in non-SUSY theories when one introduces a center-stabilizing double-trace deformation or boundary conditions \cite{Unsal:2007jx}. An underlying reason for calculability of monopole operators in non-SUSY theories is explored in \cite{unsal}.}. The reason for this is that the dynamics of the monopole-instantons can be studied systematically, and various dynamical effects of the 4D theory (such as gaugino condensation or the general Affleck-Dine-Seiberg (ADS) superpotential~\cite{Affleck:1983mk}) can be understood in terms of monopole \cite{Affleck:1982as,Aharony:1997bx} and multi-monopole \cite{multimonopoles} effects in the compactified theory. The essential property of monopoles that largely determines the structure of the induced superpotential terms is the number of fermionic zero modes in a given monopole background. The Callias index theorem \cite{Callias:1977kg,Erich} specifies the number of fermionic zero modes in different gauge group  representations for a given monopole background. However, for the compactified theory there is a twisted embedding of the monopole solution called the KK monopole. This KK monopole is obtained by performing an anti-periodic gauge transformation along the compactified circle. The effects of the KK monopole are crucial for obtaining the correct interpolation between the 4D and 3D theories. Thus it is essential to understand how the number of fermionic zero modes of the KK monopole can change.  The goal of this paper is to give a simple intuitive accounting for fermion zero modes in a KK monopole background. KK monopoles were first introduced by Lee and Yi~\cite{Lee:1997vp}, though their contribution to the superpotential was anticipated by Seiberg and Witten~\cite{Seiberg:1996nz}. Using the Nahm construction, KK monopole configurations were found explicitly in~\cite{Lee:1998bb,Davies:1999uw}. Aharony et. al ~\cite{Aharony:1997bx} already contains a brief comment on the number of fermionic zero modes. The number of fermionic zero modes was also inferred in ~\cite{Erich} using the fact that all the independent monopoles together make up a 4D instanton in the large radius limit~\cite{Lee:1998bb}. The analogs of the KK monopoles for finite temperature field theories were introduced in~\cite{Kraan:1998pm}, while an analysis of the zero modes of the finite temperature version was presented in~\cite{Bruckmann:2003ag}.  Here we give a detailed explanation of why a fermion in the fundamental representation has a zero mode in a KK monopole background only when the real mass $m$ satisfies $|m|> \frac{v}{2}$, where $v$ is the asymptotic adjoint scalar vacuum expectation value (VEV) of the monopole background. This is the exact opposite of the condition for the existence of a zero mode in the ordinary monopole background: $|m|< \frac{v}{2}$. On the other hand the condition for the existence of an adjoint fermion (gaugino) zero mode is the same for both the ordinary monopole and the KK monopole. The root cause for the unusual behavior of the fundamental zero modes is the fact that the fundamental carries a single gauge index, and hence the usual zero mode would become anti-periodic under the large gauge transformation that connects the KK monopole to the ordinary monopole. The true KK zero mode originates in a configuration that is anti-periodic around the circle before the gauge transformation is performed. Since the adjoint carries two indices, its zero mode is periodic in either case, so there is no difference in the conditions for gaugino zero modes. Our results provide an intuitive explanation of KK monopole decoupling in the limit of a large real mass: for a sufficiently large real mass the KK monopole {\it acquires} additional fundamental fermion zero modes, and as a result the KK monopole cannot correspond to a superpotential term. 

The paper is organized as follows. First we briefly review the construction of the KK monopole solution, and then remind the reader of the form of the fermionic zero modes in ordinary monopole backgrounds. Rather than relying on index theorems \cite{Callias:1977kg,Erich} we analyze the properties of the solutions of the Dirac equation in the monopole background \`a la Jackiw and Rebbi \cite{Jackiw:1975fn}, while allowing for a real mass term. Next we present our main result: the condition for the existence of fermionic zero modes in the fundamental representation in the KK monopole background. We apply our result to explain the decoupling of the effects of the KK monopole in ${\cal N}=1$ SUSY (4 supercharges) theories on ${\bf R}^3\times {\bf S}^1$. Finally we show a neat example based on the $SU(2)\times SU(2)$ theory with a bifundamental where the interplay between the fundamental fermion zero modes   of the KK monopole and the ordinary monopoles exactly reproduces the answers expected from the 4D analysis of~\cite{Intriligator:1994jr}. 

\section{BPS vs. KK Monopoles}

The fundamental BPS monopole is nothing but the usual 't Hooft-Polyakov monopole of the Georgi-Glashow model. For simplicity we will only consider the $SU(2)$ case, but all results can be readily generalized to $SU(N)$ by the embedding of $SU(2)$ subgroups. Since we have the application to SUSY gauge theories in mind we will use the holomorphic normalization of the gauge fields (where the gauge coupling $g$ appears only in the gauge kinetic terms). The explicit expression of the monopole background  is 
\begin{equation}
 A_i^a(\vec{x})=\epsilon_{ija}\,\hat x^j   \frac{f(r)}{r} \, ,\  \phi^a(\vec{x})=v \,\hat x^a h(r)
\end{equation}
where $v$ is the asymptotic adjoint scalar VEV, $r=|\vec{x}|$, and (since there is no scalar potential) the functions $h,f$ are
$f(r)=\left(1-\frac{vr}{\sinh(vr)}\right)$,  $h(r)=\left(\coth(vr)-\frac{1}{vr}\right)$, where both $f,h \to 1$ for $r\to \infty$. In the compactified Euclidean theory the scalar $\phi$ can be thought of as the fourth component of the gauge field $A_4=\phi$. In the following we use $\sigma^i$ to denote the Pauli matrices.

The construction of the KK monopole on the interval $0 \le x_4 \le 2 \pi R$ requires three steps~\cite{Lee:1998bb}. First, one replaces the asymptotic adjoint VEV $v$ with $v^\prime=1/R-v$. Then one performs a large  gauge transformation $\sim e^{-i \frac{x_4}{2 R}\sigma^3}$, which is anti-periodic along the compact $x_4$ dimension. This transformation shifts the VEV by $-1/R$. Finally, one can restore the  original VEV $v$ by a Weyl transform that takes $v\rightarrow -v$. The result of the combined  transformations takes the form \cite{Lee:1998bb}
\begin{equation}
 A_\mu=U^\dagger A_\mu(\vec{x}, v^\prime) U+i U^\dagger\partial_\mu U\,,
\end{equation}
where $A_\mu$ is the gauge field (with $A_4=\phi$) of the BPS monopole and the gauge transformation $U$ is given by \cite{Lee:1998bb}
\begin{equation}
\label{eq:largegauge}
 U=U_{h}\sigma^2 e^{-i\frac{x_4\sigma^3}{2R}}U_{h}^\dagger\,,
\end{equation}
where
\begin{equation}
 U_{h}=\frac{\sigma^3 \cosh\frac{v^\prime r}{2}+\vec{\sigma}\cdot \vec{x}\sinh\frac{v^\prime r}{2}}{\sqrt{\cosh v^\prime r+\hat x_3\sinh v^\prime r}}\,.
\end{equation}
In~(\ref{eq:largegauge}) $U_{h}$ is trivial at the origin while implementing a transformation between hedgehog and singular gauges at infinity. It is only needed to make sure that the behavior of $\phi^a$ at infinity is the same for both KK and BPS monopoles. The global transformation $\sigma^2$ implements Weyl reflection. Finally, $\sim e^{-i \frac{x_4}{2 R}\sigma^3}$ is the anti-periodic gauge transformation that flips the magnetic charge of the monopole.

\section{Zero modes of the BPS monopole}

According to the Callias index theorem a chiral fermion in the fundamental representation has one zero mode in the background of the BPS monopole. To explicitly find this zero mode we need to solve the Dirac equation \`a la Jackiw and Rebbi. We are taking the 3D theory obtained by compactifying the theory on a circle in the timelike direction, and Wick-rotated to Euclidean space with $x_4= - i x_0, A_4=- i A_0$. The equation is given by 
\begin{widetext}
\begin{eqnarray}
\left(\vec{ \nabla } \cdot \vec{\sigma}^{\, \alpha \beta}\delta^m_n + i  \vec{A}^{\,a} \cdot  \vec{\sigma}^{\, \alpha \beta} T^{am}_{\,\,\,\,n} -  \,\phi^a \,\delta^{\alpha \beta} T^{am}_{\,\,\,\,n}  - m \,\delta^{\alpha \beta} \delta^m_n \right)\psi_{\beta m}=0\, ,
\label{diracfundamental}
\end{eqnarray}
\end{widetext}
where $m$ is the real mass of the fundamental, obtained from the time component of a four-dimensional background gauge field, which weakly gauges ``baryon" number (implying that it is $SU(2)$ color invariant, hence the additional color Kronecker delta). 

We will look for solutions of the form
\begin{equation}
 \psi_{\alpha m}(\vec{x})=\left(\begin{array}{c}u\\ d\end{array}\right)=\sigma^2_{\alpha m} X(x)+(\hat x^a\sigma^a \sigma^2)_{\alpha m} Y(x)\,.
\end{equation}
With this ansatz, the zero mode must satisfy the equations
\begin{eqnarray}
Y'+\frac{2-f}{r} Y + \frac{v}{2} h Y = m X \nonumber \\
X'+\frac{f}{r} X +\frac{v}{2} h X = mY 
\end{eqnarray}
For $m=0$ the two equations are decoupled and can be integrated. The requirement that the solution is normalizable implies that $Y=0$. The single zero mode in this case is then given by \cite{Jackiw:1975fn}
\begin{equation}
X(r) = C e^{-\int_0^r \left( \frac{v}{2} h(r') +\frac{f(r')}{r'}\right) dr' }
\end{equation}
which is normalizable since $h(r)\to 1,  \frac{f(r)}{r}\to 0$ as $r\to \infty$.  For the case with a real mass $m$ we need to solve the second order differential equation
\begin{equation}
\left( \frac{d}{dr} + \frac{2-f}{r} +\frac{v}{2}h \right)\left( \frac{d}{dr} +\frac{f}{r} +\frac{v}{2} h\right) X=m^2 X\ .
\label{eq:secondorder}
\end{equation}
When $X$ is normalizable the asymptotic behavior is $X \sim e^{-\lambda r}$, with $\lambda >0$. Eq.~(\ref{eq:secondorder}) then implies $(\frac{v}{2}-\lambda )^2 =m^2$. There is a positive solution for $\lambda$ provided that
\begin{equation}
-\frac{v}{2} < m < \frac{v}{2}
\end{equation}
which exactly agrees with the Callias index theorem \cite{Callias:1977kg,Erich}. 

\section{Zero modes of the KK monopole}

It is well-known that for a vanishing real mass, the KK-monopole does not have a normalizable zero mode for fermions in the fundamental representation. Next we explain why this is so and show that for sufficiently large real masses normalizable zero modes do exist. The essential physics insight is the fact that a fundamental fermion behaves differently under an anti-periodic gauge transformation than an adjoint fermion due to the fact that it carries only a single $SU(2)$ index. A large gauge transformation on adjoints introduces a periodic dependence on the coordinate along the compactified circle. However fields in the fundamental pick up an additional sign and thus would become anti-periodic. 
Thus, the expectation is that while gaugino zero modes in the KK monopole background exist and can be obtained by a large gauge transformation (\ref{eq:largegauge}), the fundamental fermion has no zero modes. However, a careful examination of the properties of the anti-periodic solution suggests that a new, twisted, zero mode of the fundamental fermion exists for sufficiently large mass. Since the large gauge transformation introduces an additional anti-periodic phase for the fundamental fermion, we need to look for an {\it anti-periodic} solution to the Dirac equation in the BPS monopole background, with the VEV shifted to $v'=\frac{1}{R}-v$. Thus we look for an ansatz of the form
\begin{equation}
 \psi(x,x_4)=e^{\pm\frac{ ix_4}{2R}} \psi(x)\, 
\label{eq:antiperiodic}
\end{equation}
which is anti-periodic for both sign choices.
For this ansatz, the $\partial_4$ derivative shifts the fermion mass by $\pm 1/(2R)$; thus the 3-dimensional part of the Dirac equation has an effective mass 
\begin{equation}
 m_{\textrm{eff}}=m \mp \frac{1}{2R}\,.
\end{equation}
The condition for the existence of a normalizable zero mode solution $|m_{\textrm{eff}}| < \frac{v'}{2}$ is translated to $|m\mp \frac{1}{2R}| < \frac{1}{2R} -\frac{v}{2}$, which can be satisfied provided 
\begin{equation}
m > \frac{v}{2} \ \ {\rm or} \ \ m < -\frac{v}{2} \ .
\end{equation}
While this solution is anti-periodic and not physical in the BPS monopole background, after the application of the large gauge transformation it  becomes periodic and provides the proper zero mode in the KK monopole background. Note that the final form of the solution will be 
\begin{equation}
 \psi(x,x_4)=\left(\begin{array}{c} e^{in \frac{x_4}{R}}u\\ e^{i(n+1)\frac{x_4}{R}} d\end{array}\right)\,,
\end{equation}
where $n=0$ corresponds to the choice of the + sign in (\ref{eq:antiperiodic}) and $n=-1$ to the - sign. 
It is easy to generalize this result to the case of a fundamental representation of $SU(N)$. In this case there is a monopole solution for each simple root $\alpha_i$, where $i=1,\ldots , N-1$.  Writing an adjoint VEV as ${\rm diag} (v_1, v_2, \ldots , v_N)$ with $\sum_i v_i=0$ and $v_i > v_{i+1}$, the fundamental zero mode lives on the  monopole associated with $\alpha_i$ if $v_{i+1}/2 < m < v_{i}/2$, and on the KK monopole for $m> v_1/2$ or $m< v_{N}/2$.  
We note finally that the KK monopole acquires a zero mode exactly as the zero mode disappears from the BPS monopole. This means that the total number of zero modes in $N$-monopole backgrounds is independent of the real mass and always matches the number of fermionic zero modes of the 4D instanton.

\section{KK monopole decoupling}

The physical importance of KK monopole zero modes becomes obvious if we consider\footnote{Another interesting case was recently studied in \cite{Cherman:2016hcd} where KK monopoles were shown to play a role in chiral symmetry breaking.} gauge theories with ${\cal N}=1$ SUSY  (4 supercharges) on $\bf{R}^3\times \bf{S}^1$. This theory can be used to interpolate between the 4D theory (taking the radius of the circle $R\to \infty$) and the 3D theory (by taking $R$ very small). 
However, the $R\to 0$ limit is not sufficient to obtain a truly 3D theory since, as noted in \cite{Aharony:1997bx}, rather than reproducing a truly 3D SUSY gauge theory, one arrives at the theory deformed by a tree level superpotential $\eta Y$, where $Y$ is the KK  monopole operator parametrizing the Coulomb branch. While $\eta$ vanishes in the $R\to 0$ limit, the presence of such an operator is problematic for 3D duality since KK monopole operators appear on both sides of the duality and force the duality scale to zero. The appearance of KK monopole zero modes resolves the problem and allows for the derivation of 3D dualities. Generically, there are several monopole operators $Y_i$ corresponding to the simple roots of the gauge group. Semiclassically these monopole operators are given by $Y_i \sim e^{4\pi (v_{i+1}-v_{i})/g_3^2}$, where the $v_i$'s are the adjoint VEVs (which can be promoted to chiral superfields), and $g_3$ is the 3D gauge coupling. Whenever there are exactly two fermionic zero modes  for a BPS monopole a superpotential term of the form $1/Y_i$  is generated. On the other hand, the action of a KK monopole is proportional to $4\pi [1/R- (v_i-v_N)]/g_3^2$, thus giving a contribution $\sim e^{-4\pi /R g_3^2} \prod_i Y_i$. The first factor $\eta = e^{-4\pi /R g_3^2}$ can be thought of as the analog of the 4D instanton factor $\Lambda_4^b$ if one matches the 3D and 4D gauge couplings: $2 \pi R \,g_3^2 = g_4^2$. It is conventional to define $Y \equiv \prod_i Y_i$.  It is the presence of the additional $\eta Y$ term upon compactification that enforces some of the 4D properties on the 3D theory, and therefore it is essential that one properly decouple this term in order to arrive at a true 3D theory without deformations. 

The proposal of~\cite{Aharony:2013dha} was to add a large real mass to one of the quark flavors. Naively one could think that decoupling a single flavor would just change the $\eta Y$ term of the KK monopole to an effective $\tilde{\eta} \tilde{Y}$ of the theory with the number of quark flavors reduced by one. Aharony {\em et} al. however argued \cite{Aharony:2013dha} that a large real mass for a single flavor completely removes the $\eta Y$ term: an effective $\tilde{\eta}\tilde{Y}$ would necessarily depend upon the real mass of the flavor that was decoupled, but the real mass can not appear in a holomorphic quantity and hence there can be no $\tilde{\eta} \tilde{Y}$ in the effective superpotential. However the dynamical origin of the KK monopole decoupling from the superpotential is not intuitively clear from this argument. Indeed, the KK monopole itself obviously still exists even when one flavor becomes heavy. Thus it can only decouple if the number of fermion zero modes changes. Since gaugino zero modes exist independently of the real mass for the fundamental flavor, the decoupling would require an appearance of new zero modes and this is precisely what we found. Once the real mass is raised above $v/2$ the KK monopole no longer contributes to a chiral fermionic two-point correlation function and thus does not generate a superpotential term. This provides a dynamical explanation for the decoupling of the effects of the KK monopole and hence the explanation of how the undeformed 3D theory is approached in this limit. 

\section{$SU(2)\times SU(2)$ with a bifundamental}
In this section we illustrate the importance of KK monopole zero modes by considering a supersymmetric  $SU(2)\times SU(2)$ theory with four supercharges and  matter $Q$ in the bifundamental representation.\footnote{A non-supersymmetric theory with similar matter content has been analyzed in~\cite{MithatMisha}.} The superpotential of this theory was found to be \cite{Intriligator:1994jr}:
\begin{equation}
\label{eq:su2squared}
 W=\frac{\left(\Lambda_1^{5/2}\pm \Lambda_2^{5/2}\right)^2}{Q^2}\,,
\end{equation}
where $Q^2$ is a gauge invariant meson constructed out of the bifundamental. In 4D the origin of this superpotential can be understood as follows: both $SU(2)$ factors have the right number of flavors to produce an instanton generated ADS superpotential term. Along the Higgs branch parametrized by the meson VEV $Q^2$ the gauge group is broken to a diagonal $SU(2)_D$ and there are no charged light fields remaining. Gaugino condensation in the low-energy gauge group contributes another $\pm 2 \Lambda_D^3$ term to the superpotential. The superpotential (\ref{eq:su2squared}) arises as a combination of these three effects.

Let us now consider the dynamics of this model on $\bf{R}^3\times \bf{S}^1$ and then recover 4D physics by taking $R\to \infty$. The classical moduli space contains a Coulomb branch parametrized by the adjoint VEVs $v_1$, $v_2$ as well as a Higgs branch parametrized by the squark VEV $Q$. The adjoint VEVs break $SU(2)_1\times SU(2)_2$ to $U(1)_1\times U(1)_2$, while the  squark VEV breaks $SU(2)_1\times SU(2)_2$ to the diagonal subgroup $SU(2)_D$. For concreteness we will assume that $v_1>v_2\gg Q$. It is important to note that from the point of view of the $SU(2)_1$ dynamics, the $v_2$ VEV serves as a real mass term for the $SU(2)_1$ doublets. Similarly, the $v_1$ VEV serves as a real mass for the $SU(2)_2$ doublets. We can see that the BPS monopole of $SU(2)_1$ and the KK monopole of $SU(2)_2$ have two gaugino and two quark zero modes, while the KK monopole of $SU(2)_1$ and the BPS monopole of $SU(2)_2$ only have two gaugino zero modes. 

At first sight one might conclude that there is a superpotential contribution from the first KK monopole and the second BPS monopole, but the actual dynamics is somewhat more intricate. Since there is a squark VEV turned on, it will break the two $U(1)$'s to the diagonal, $U(1)_1\times U(1)_2 \to U(1)_D$, and monopoles which carry the broken $U(1)$ charge are confined. Thus only multi-monopole configurations neutral under the broken $U(1)$ will contribute to the superpotential. There are four such multi-monopole configurations made out of two confined monopoles:  the first BPS and the first KK monopoles,  the second BPS and the second KK monopoles, the two BPS monopoles, and the two KK monopoles. While these multi-monopole solutions each have several zero modes, some of them can be soaked up using the squark VEV each eventually yielding contributions to the superpotential. Which zero modes are lifted is determined by the pattern of $U(1)$ breaking since the corresponding gaugino gets a mass with a quark via the squark VEV as required by SUSY.

 For example, the double monopole made of the first BPS and first KK monopoles generates the expected ADS term in the superpotential,
\begin{equation}
 W_1=\frac{\eta_1}{Q^2}\,.
\label{eq:W1}
\end{equation}
 In fact, this two monopole configuration is equivalent \cite{Lee:1998bb} to a periodic instanton on $\bf{R}^3\times \bf{S}^1$. Similarly, the configuration  made up of the second BPS and KK monopoles leads to the instanton-generated ADS superpotential in $SU(2)_2$ even though the distribution of fermion zero modes between BPS and KK monopoles is different here:
\begin{equation}
 W_2=\frac{\eta_2}{Q^2}\,.
\label{eq:W2}
\end{equation}

Finally, the configurations with the two BPS and the two KK monopoles (which act as the monopoles of $SU(2)_D$) produce the  superpotential
\begin{equation}
 W_{1,2}=\frac{1}{Q^2}\left(\eta_1\eta_2 Y_1Y_2+\frac{1}{Y_1Y_2}\right)\,.
\label{eq:confineddiagonal}
\end{equation}
 Solving the equations of motion for the composite monopole $Y_1Y_2$ we find that  (\ref{eq:confineddiagonal}) will contribute $\pm \frac{2 \sqrt{\eta_1 \eta_2}}{Q^2}$, which together with (\ref{eq:W1}) and (\ref{eq:W2}) results in the correct superpotential (\ref{eq:su2squared}).

\section{Conclusions}

Fermionic zero modes of monopoles largely determine the structure of the dynamical monopole-induced effects in supersymmetric theories. We have found the condition for the existence of fermionic zero modes in the fundamental representation in the KK monopole background, and showed that such zero modes will be present for a sufficiently large real mass term. This explains the previously mysterious decoupling of the effects of KK monopoles in theories with four supercharges in the presence of a large real mass, which allows one to explore the dynamics of a truly 3D theory. We have applied our results to the $SU(2)\times SU(2)$ model with a bifundamental and shown that the terms attributed to gaugino condensation in 4D originate from multi-monopole terms in the 3D theory.

\section{Acknowledgements}

We thank Ken Intriligator, Mario Martone, Kimyeong Lee, and Mithat Unsal for helpful discussions and comments on the manuscript. C.C. is supported in part by NSF grant PHY-1316222. Y.S. is supported in part by NSF grant PHY-1620638. J.T is supported in part by DOE grant DE-SC0009999. We thank the hospitality and partial support of the Mainz Institute for Theoretical Physics (MITP) during the completion of this work.


\begin{thebibliography}{99}

\bibitem{Seiberg:1994rs}
  N.~Seiberg and E.~Witten,
  ``Electric-magnetic duality, monopole condensation, and confinement in ${\mathcal N}=2$ supersymmetric Yang-Mills theory,''
Nucl.\ Phys.\ B {\bf 426} (1994) 19 \arXivold{hep-th/9407087}.

\bibitem{Seiberg:1996nz} 
  N.~Seiberg and E.~Witten,
  ``Gauge dynamics and compactification to three-dimensions,''
In {\em Saclay 1996, The mathematical beauty of physics} 333-366
\arXivold{hep-th/9607163}.


\bibitem{Aharony:1997bx}
  O.~Aharony, A.~Hanany, K.~A.~Intriligator, N.~Seiberg and M.~J.~Strassler,
  ``Aspects of ${\mathcal N}=2$ supersymmetric gauge theories in three-dimensions,''
Nucl.\ Phys.\ B {\bf 499} (1997) 67
\arXivold{hep-th/9703110}.


\bibitem{deBoer}
  J.~de Boer, K.~Hori, H.~Ooguri and Y.~Oz,
  ``Mirror symmetry in three-dimensional gauge theories, quivers and D-branes,''
Nucl.\ Phys.\ B {\bf 493} (1997) 101
\arXivold{hep-th/9611063}.

\bibitem{MithatEric}
  E.~Poppitz and M.~Unsal,
  ``Seiberg-Witten and 'Polyakov-like' magnetic bion confinements are continuously connected,''
JHEP {\bf 1107} (2011) 082
\arXiv{1105.3969}{hep-th}.


\bibitem{Aharony:2013dha}
  O.~Aharony, S.~S.~Razamat, N.~Seiberg and B.~Willett,
  ``3d dualities from 4d dualities,''
JHEP {\bf 1307} (2013) 149
\arXiv{1305.3924}{hep-th}.


\bibitem{Csaki:2014cwa} 
  C.~Cs\'aki, M.~Martone, Y.~Shirman, P.~Tanedo and J.~Terning,
  ``Dynamics of 3D SUSY Gauge Theories with Antisymmetric Matter,''
JHEP {\bf 1408} (2014) 141
\arXiv{1406.6684}{hep-th};
  A.~Amariti, C.~Cs\'aki, M.~Martone and N.~R.~L.~Lorier,
  ``From 4D to 3D chiral theories: Dressing the monopoles,''
Phys.\ Rev.\ D {\bf 93} (2016) 105027
\arXiv{1506.01017}{hep-th}.

\bibitem{Csaki:2011nw} 
  C.~Csaki, D.~Curtin, V.~Rentala, Y.~Shirman and J.~Terning,
 ``Supersymmetry Breaking Triggered by Monopoles,''
  Phys.\ Rev.\ D {\bf 85}, 045014 (2012),
  \arXiv{1108.4415}{hep-th}.

\bibitem{Unsal:2007jx} 
  M.~Unsal,
  ``Magnetic bion condensation: A New mechanism of confinement and mass gap in four dimensions,''
  Phys.\ Rev.\ D {\bf 80} (2009) 065001
  \arXiv{0709.3269}{hep-th};
  M.~Unsal and L.~G.~Yaffe,
  ``Center-stabilized Yang-Mills theory: Confinement and large $N$ volume independence,''
  Phys.\ Rev.\ D {\bf 78} (2008) 065035
  \arXiv{0803.0344}{hep-th}.

\bibitem{unsal}
Z.~Komargodski,  T.~Sulejmanpasic, and M.~Unsal, work in progress.



\bibitem{Affleck:1983mk}
  I.~Affleck, M.~Dine and N.~Seiberg,
  ``Dynamical Supersymmetry Breaking in Supersymmetric QCD,''
\href{http://dx.doi.org/10.1016/0550-3213(84)90058-0}{Nucl.\ Phys.\ B {\bf 241} (1984) 493}.


\bibitem{Affleck:1982as} 
  I.~Affleck, J.~A.~Harvey and E.~Witten,
  ``Instantons and (Super)Symmetry Breaking in (2+1)-Dimensions,''
\href{http://dx.doi.org/10.1016/0550-3213(82)90277-2}{Nucl.\ Phys.\ B {\bf 206} (1982) 413}.



\bibitem{multimonopoles}
C.~Cs\'aki, M.~Martone, Y.~Shirman and J.~Terning,
  ``Pre-ADS Superpotential from Confined Monopoles,''
  \arXiv{1711.11048}{hep-th}.


\bibitem{Callias:1977kg} 
  C.~Callias,
``Index Theorems on Open Spaces,''
 \href{http://dx.doi.org/10.1007/BF01202525}{Commun.\ Math.\ Phys.\  {\bf 62}, 213 (1978)};

\bibitem{Erich}
  T.~M.~W.~Nye and M.~A.~Singer,
 ``An $L^2$ index theorem for Dirac operators on ${\bf S}^1 \times {\bf R}^3$,''
\arXivold{math/0009144};
  E.~Poppitz and M.~Unsal,
  ``Index theorem for topological excitations on ${\bf R}^3 \times {\bf S}^1$ and Chern-Simons theory,''
JHEP {\bf 0903} (2009) 027
\arXiv{0812.2085}{hep-th}.

\bibitem{Lee:1997vp}
  K.~M.~Lee and P.~Yi,
  ``Monopoles and instantons on partially compactified D-branes,''
Phys.\ Rev.\ D {\bf 56} (1997) 3711
\arXivold{hep-th/9702107};


\bibitem{Lee:1998bb}
  K.~M.~Lee,
  ``Instantons and magnetic monopoles on ${\bf R}^3 \times {\bf S}^1$ with arbitrary simple gauge groups,''
Phys.\ Lett.\ B {\bf 426} (1998) 323
\arXivold{hep-th/9802012};
  K.~M.~Lee and C.~h.~Lu,
  ``SU(2) calorons and magnetic monopoles,''
Phys.\ Rev.\ D {\bf 58} (1998) 025011
\arXivold{hep-th/9802108}.




\bibitem{Davies:1999uw}
  N.~M.~Davies, T.~J.~Hollowood, V.~V.~Khoze and M.~P.~Mattis,
  ``Gluino condensate and magnetic monopoles in supersymmetric gluodynamics,''
Nucl.\ Phys.\ B {\bf 559} (1999) 123
\arXivold{hep-th/9905015};
  N.~M.~Davies, T.~J.~Hollowood and V.~V.~Khoze,
  ``Monopoles, affine algebras and the gluino condensate,''
J.\ Math.\ Phys.\  {\bf 44} (2003) 3640
\arXivold{hep-th/0006011}.






\bibitem{Kraan:1998pm}
  T.~C.~Kraan and P.~van Baal,
  ``Periodic instantons with nontrivial holonomy,''
Nucl.\ Phys.\ B {\bf 533} (1998) 627
\arXivold{hep-th/9805168}.




\bibitem{Bruckmann:2003ag} 
  F.~Bruckmann, D.~Nogradi and P.~van Baal,
  ``Constituent monopoles through the eyes of fermion zero modes,''
Nucl.\ Phys.\ B {\bf 666} (2003) 197
\arXivold{hep-th/0305063}.



\bibitem{Jackiw:1975fn}
  R.~Jackiw and C.~Rebbi,
  ``Solitons with Fermion Number 1/2,"
    \href{http://journals.aps.org/prd/pdf/10.1103/PhysRevD.13.3398}{Phys.\ Rev.\ D {\bf 13} (1976) 3398.}

\bibitem{Cherman:2016hcd} 
  A.~Cherman, T.~Sch\"afer and M.~\"Unsal,
  Phys.\ Rev.\ Lett.\  {\bf 117}  (2016) 081601
  \arXiv{1604.06108}{hep-th}.

\bibitem{MithatMisha}
  M.~Shifman and M.~Unsal,
  ``On Yang-Mills Theories with Chiral Matter at Strong Coupling,''
Phys.\ Rev.\ D {\bf 79} (2009) 105010
\arXiv{0808.2485}{hep-th}.



\bibitem{Intriligator:1994jr}
  K.~A.~Intriligator, R.~G.~Leigh and N.~Seiberg,
  ``Exact superpotentials in four-dimensions,''
Phys.\ Rev.\ D {\bf 50} (1994) 1092
\arXivold{hep-th/9403198}.


\end{thebibliography}
\end{document}